\documentclass[12pt,a4paper,final]{iopart}

\usepackage{iopams}  
\usepackage{graphicx}
\usepackage[breaklinks=true,colorlinks=true,linkcolor=blue,urlcolor=blue,citecolor=blue]{hyperref}
\expandafter\let\csname equation*\endcsname\relax
\expandafter\let\csname endequation*\endcsname\relax
\usepackage{amsmath}
\begin{document}

\title{Autoionizing States driven by Stochastic Electromagnetic Fields}

\author{G. Mouloudakis$^1$ and P. Lambropoulos$^{1,2}$}

\address{${^1}$Department of Physics, University of Crete, P.O. Box 2208, GR-71003 Heraklion, Crete, Greece}

\address{${^2}$Institute of Electronic Structure and Laser, FORTH, P.O.Box 1527, GR-71110 Heraklion, Greece}
\ead{gmouloudakis@physics.uoc.gr }

\begin{abstract}
We have examined the profile of an isolated autoionizing resonance driven by a pulse of short duration and moderately strong field. The analysis has been based on stochastic differential equations governing the time evolution of the density matrix under a stochastic field.  Having focused our quantitative analysis on the
$2s2p({}^1P)$ resonance of Helium, we have investigated the role of field fluctuations and of the duration of the pulse. We report surprisingly strong distortion of the profile, even for peak intensity below the strong field limit. Our results demonstrate the intricate connection between intensity and pulse duration, with the latter appearing to be the determining influence, even for a seemingly short pulse of 50 fs. Further effects that would arise under much shorter pulses are discussed.        
\end{abstract}

\vspace{2pc}

Autoionizing (AI) states, also referred to as resonances, in atoms and molecules belong to a rich field of Atomic Molecular and Optical (AMO) physics, representing a paradigm of discrete states embedded in continua.  Typically, they can be excited either by photoabsorption or by collisions. The literature on the subject is vast, but luckily a relatively recent review \cite{ref1}  provides an interesting guide to the origins of the field, as well as a substantial collection of references to both theoretical and experimental work. The simplest case in point is represented by a so-called "isolated AI resonance", which means that the width of its excitation profile is much smaller than the energy distance from the nearest AI resonance. A textbook example of an isolated resonance is provided by the doubly excited $2s2p({}^1P)$ state of Helium, which has been studied in exhaustive detail, both theoretically and experimentally, over the last 60 or so years \cite{ref1}. The field has more recently been enriched with fascinating details on the temporal evolution of AI through the exploitation of coherent few cycle pulses \cite{ref2,ref3,ref4}. 

Traditional photoabsorption studies of AI resonances have been carried out mainly by means of synchrotron radiation, where weak, practically monochromatic radiation excites the resonance. The most easily observed quantities, as a function of the photon energy scanned around the resonance, can be either the amount of ionization  or photoabsorption; the number of photons missing from the beam. A widely used expression for the profile of ionization as a function of photon energy is given by $P = {{(q + \varepsilon )}^2}/({{1 +{ \varepsilon}^2}})$, where $q$ is the so-called asymmetry parameter and $\varepsilon$ the detuning of the radiation from resonance divided by the autoionization half-width $\Gamma /2$. Thus both $q$ and $\varepsilon$ are dimensionless quantities, with $q$ characterizing the degree of interference between the direct transition to the continuum and the indirect transition via the resonance, which depends on the atomic system and the particular AI resonance. For $q \gg 1$ the  profile becomes a Lorentzian, while for smaller values of $q$, it becomes asymmetric, often referred to as Fano profile \cite{ref5}. For excitation under weak (the exact meaning of which is defined below), monochromatic radiation, this is in fact the profile that will be observed. It should be kept in mind that the above expression for $P$ represents the transition probability per unit time, as given by Fermi's golden rule in lowest order perturbation theory (LOPT). It simply expresses the rate with which the ground state is depopulated as a function of the detuning around the resonance.

A brief excursion into theory, clarifying the notions of weak and strong field, is in order at this point. In the simplest formulation, an isolated AI resonance can be cast in terms of a discrete state embedded in the continuum into which it decays via intra-atomic (Coulomb) interaction. The discrete state, as well as the continuum, are also connected to the initial (typically ground) state via an electric dipole transition. Those two transition amplitudes (paths) leading to the same energy of the continuum interfere, giving rise to the asymmetric absorption profile, as long as one of the amplitudes does not dominate the other, which corresponds to a relatively small value of $q$; say below 10.  Although the formulation can begin with a discreet state, a continuum and two interactions, namely the electric dipole coupling $D$ and the intra-atomic coupling $V$, even in the simplest treatment \cite{ref5}, the discrete state is modified through an admixture of the continuum, mediated by $V$. Hereafter, it is the dipole coupling of the initial to that modified discreet state that is meant by coupling to the discreet part of the resonance. The product of that dipole matrix element and the electric field amplitude $\mathcal{E}(t)$ divided by 2, represents the Rabi frequency denoted by $\Omega $. Note that the electric field is defined as $E(t) = \frac{1}{2}[\mathcal{E} (t){e^{i\omega t}} + c.c.]$. The time dependence in $\mathcal{E}(t)$ is meant to indicate the deterministic time dependence due to the pulse, as well as the stochastic time dependence due to the fluctuations of the field. The rate of the direct transition from the initial state to the continuum is given by $\gamma  = \sigma F$ where $\sigma$ is the photoionization cross section to the smooth continuum, away from the resonance and $F$ the photon flux in $photons/c{m^2}s$. The above four parameters are related through the equation $4{\Omega ^2} = {q^2}\gamma \Gamma $. Their values must be obtained through an elaborate atomic structure calculation, which has been done in numerous papers \cite{ref6,ref7,ref8}. For the particular case of He(2s2p) AI resonance they are: $q =  - 2.75$, $\Gamma  = 1.37 \times {10^{ - 3}}(a.u.)$, $\sigma  = {10^{ - 18}}c{m^2}$, and $\Omega  = 0.025\frac{{\rm \mathcal{E}(t)}}{2}(a.u.)$. In consistency with these values, the ionization rate directly into the smooth continuum is given by $\gamma  = 0.1175 \times {\rm I}(a.u.)$, where $I$ is the intensity. In the Rotating Wave Approximation (RWA) \cite{ref9}, the relation between the intensity and our definition of the electric field is $I = \frac{1}{2}\mathcal{E}(t)^2$. Clearly ${\Omega ^2}$ and $\gamma$ are proportional to the radiation intensity, which cancels out in the equation that constrains the four parameters. With all of the atomic parameters in place, we can now define the criterion for strong versus weak field. The field will
hereafter be considered strong when $\Omega$ is larger
than $\Gamma$ (${\Omega}>{\Gamma}$). Since the fields assumed in this work are time dependent, so is $\Omega$. Following the standard practice in laser-atom interactions, it will be the value of $\Omega$ at peak intensity that will be understood in the context of the above inequality. 

The dynamics of the evolution of an AI resonance under a pulse of significant intensity, such as delivered by a short wavelength Free Electron Laser (FEL), requires a formulation in terms of the density matrix ${\rho}(t)$ \cite{ref9,ref10}, because it allows in addition the inclusion of the stochastic properties of the radiation, inherent in the FEL. Using the parameters defined above, the differential equations governing the time dependence of the slowly varying part ${\sigma}(t)$ of the density matrix in the Rotating Wave Approximation (RWA) are:

\begin{equation}
{\partial _t}{\sigma _{11}}(t) =  - \gamma (t){\sigma _{11}}(t) + 2Im\left\{ {\Omega (t)\left( {1 - \frac{i}{q}} \right){\sigma _{21}}(t)} \right\}
\end{equation}
\begin{equation}
{\partial _t}{\sigma _{22}}(t) =  - \Gamma {\sigma _{22}}(t) - 2Im\left\{ {\Omega (t)\left( {1 + \frac{i}{q}} \right){\sigma _{21}}(t)} \right\}
\end{equation}
\begin{equation}
\left[ {{\partial _t} - i\Delta  + \frac{1}{2}\left( {\gamma (t) + \Gamma } \right)} \right]{\sigma _{21}}(t) =  - i\Omega (t)\left( {1 - \frac{i}{q}} \right){\sigma _{11}}(t) + i\Omega (t)\left( {1 + \frac{i}{q}} \right){\sigma _{22}}(t)
\end{equation}

where we have introduced the slowly varying amplitudes ${\sigma _{ij}}(t)$, obeying ${\rho _{ii}}(t) = {\sigma _{ii}}(t)$, for $i = 1,2$ and ${\rho _{21}}(t) = {\sigma _{21}}(t)exp[i\omega t]$. The detuning $\Delta$ of the photon frequency $\omega$ from resonance is defined by $\Delta  \equiv \omega  - {\omega _{21}}$, with 
$\omega_{21}$ denoting the energy difference between the ground state $\left| 1 \right\rangle $ and the discrete part of the resonance $\left| 2 \right\rangle $. The left side of equation (3) may in general contain additional coherence (off-diagonal) relaxation constants which are of no relevance in the case of monochromatic field. However, a coherence relaxation constant will appear when we introduce field fluctuations.

The above set of equations provide the fundamental and versatile tool enabling the exploration of aspects beyond the simple weak field, Fermi golden rule excitation of the resonance.  The problem of strong coupling was addressed as early as 1980 and several of its aspects have since been belabored in theoretical papers \cite{ref11,ref12}. The chief point concerns the profile of an AI resonance driven by a field strong in the sense defined above. Of particular interest in that regime is the problem of double resonance, coupling strongly two AI states, where an AC Stark splitting is expected \cite{ref11,ref12,ref13}. In that case the field must be such that the effective Rabi frequency
coupling the two resonances is larger than the widths of both. Experimental investigation of that aspect has up to now been rather limited \cite{ref14,ref15}. The theory of double resonance has in addition been extended to the case of strong coupling between triply excited hollow states \cite{ref16,ref17,ref18}, whose experimental exploration requires high intensity sources of photon energies in the range of 100 eV. We shall not be concerned with those issues in this paper.

Given that the FEL radiation is subject to stochastic fluctuations \cite{ref19}, akin to those of a thermal (chaotic) state \cite{ref20,ref21}, the treatment must incorporate that property; which requires a generalization of the above density matrix equations. Sidestepping the complete mathematical derivation \cite{ref22,ref23}, we outline here the basic steps.
Note first that, owing to the field fluctuations, the parameters involving the field, namely $\Omega$ and 
$\gamma$, are also fluctuating quantities. Therefore the differential equations for the time evolution of the density matrix involve time-dependent coefficients undergoing stochastic fluctuations. This causes the matrix elements of the density matrix to undergo stochastic fluctuations, which means that we now have stochastic differential equations. As a result, the observables, such as ion yields and /or photon absorption which are obtained from these differential equations must be averaged over the stochastic fluctuations. The mathematical problem comes in the process of converting the differential equations to stochastic differential equations, which govern the evolution of the averaged quantities. 

Depending on the nature of the stochastic field, there may exist a model lending itself to an exact analytic averaging procedure \cite{ref22,ref23}. Alternatively one may always resort to numerical simulation of the stochastic averages \cite{ref21}. In fact the stochastic properties of the FEL do in general demand numerical simulation. There are however situations in which, even for a system driven by radiation with the properties of FEL, analytical results do provide an excellent approximation, as discussed below.

Briefly, the steps leading to such an approximation are as follows:

We solve equation (3) formally for ${\sigma _{21}}(t)$  and substitute the result into equations (1) and (2). Denoting the stochastic averages by angular brackets, the resulting equations are:

\begin{equation}
{\partial _t}\left\langle \sigma _{11}(t) \right\rangle = 
- \left\langle \gamma (t) \sigma _{11} (t) \right\rangle  
+ 2{\rm Im}  \left\{ \left (1 - \frac{i}{q} \right )
\int_0^t -i \left (1 - \frac{i}{q} \right )  
\left\langle {\Omega (t)\Omega (t') \sigma _{11}(t')} \right\rangle 
e^{ - \kappa (t - t')} dt' \right. $$
$$ +\left. \left (1 - \frac{i}{q} \right ) \int_0^t i\left (1 + \frac{i}{q} \right )
\left\langle \Omega (t)\Omega (t')\sigma _{22} (t') \right\rangle 
e^{ - \kappa (t - t')}dt'  \right\} 
\end{equation}

\begin{equation}
{\partial _t}\left\langle \sigma _{22} (t) \right\rangle =  
 - \Gamma \left\langle \sigma _{22}(t) \right\rangle  
 - 2{\rm Im} \left\{ \left (1 + \frac{i}{q}\right )
\int_0^t -i \left (1 - \frac{i}{q} \right )  
\left\langle {\Omega (t)\Omega (t') \sigma _{11}(t')} \right\rangle 
e^{ - \kappa (t - t')} dt' \right. $$
$$ +\left. \left (1 + \frac{i}{q} \right ) \int_0^t i\left (1 + \frac{i}{q} \right )
\left\langle \Omega (t)\Omega (t')\sigma _{22} (t') \right\rangle 
e^{ - \kappa (t - t')}dt'  \right\} 
\end{equation}
where to compress notation we have defined $\kappa  \equiv  - i\Delta  + \frac{1}{2}(\gamma  + \Gamma )$.

Equations (4) and (5) involve atom-field correlation functions of the form $\left\langle {\Omega (t)\Omega (t'){\sigma _{ii}}(t')} \right\rangle $, $i = 1,2$. In general, such correlation functions cannot be evaluated without taking into account the specific form of the fluctuations of the field. As an approximation valid under conditions discussed below, one can decorrelate the atomic-field dynamics by taking $\left\langle {\Omega (t)\Omega (t'){\sigma _{ii}}(t')} \right\rangle  = \left\langle {\Omega (t)\Omega (t')} \right\rangle \left\langle {{\sigma _{ii}}(t')} \right\rangle $. This will be referred to hereafter as the decorrelation approximation (DA).

The resulting stochastic differential equations now are:  
\begin{equation}
{\partial _t}\left\langle {{\sigma _{11}}(t)} \right\rangle  =  -\left\langle \gamma (t)\right\rangle\left\langle {{\sigma _{11}}(t)} \right\rangle  + d_{12}^2I(t){\mathop{\rm Im}\nolimits} \left\{ \left( {1 - \frac{i}{q}} \right)\int\limits_0^t  - i\left( {1 - \frac{i}{q}} \right)\left\langle {{\sigma _{11}}(t')} \right\rangle {e^{ - \tilde \kappa (t - t')}}dt' \right. $$
$$+\left. \left( {1 - \frac{i}{q}} \right)\int\limits_0^t {i\left( {1 + \frac{i}{q}} \right)\left\langle {{\sigma _{22}}(t')} \right\rangle {e^{ - \tilde \kappa (t - t')}}dt'}  \right\}
\end{equation}
\begin{equation}
{\partial _t}\left\langle {{\sigma _{22}}(t)} \right\rangle  =  - \Gamma \left\langle {{\sigma _{22}}(t)} \right\rangle  - d_{12}^2I(t){\mathop{\rm Im}\nolimits} \left\{ \left( {1 + \frac{i}{q}} \right)\int\limits_0^t  - i\left( {1 - \frac{i}{q}} \right)\left\langle {{\sigma _{11}}(t')} \right\rangle {e^{ - \tilde \kappa (t - t')}}dt' \right. $$
$$+\left. \left( {1 + \frac{i}{q}} \right)\int\limits_0^t {i\left( {1 + \frac{i}{q}} \right)\left\langle {{\sigma _{22}}(t')} \right\rangle {e^{ - \tilde \kappa (t - t')}}dt'}  \right\}
\end{equation}
where ${d_{12}}$ is the dipole matrix element between the two states and $\tilde \kappa  \equiv \kappa  + \frac{1}{2}{\gamma _L} =  - i\Delta  + \frac{1}{2}(\gamma  + \Gamma  + {\gamma _L})$. The quantity ${{\gamma _L}}$ represents the laser bandwidth arising from the first order correlation function $\left\langle {{\Omega ^*}(t)\Omega (t')} \right\rangle  = \left\langle {{{\left| {\Omega (t)} \right|}^2}} \right\rangle \exp \left[ { - \frac{1}{2}{\gamma _L}\left| {t - t'} \right|} \right]$, assuming that the field is Markovian (note that ${\Omega (t)}$ is real). The decorrelation of the variables $\gamma (t)$ and ${{\sigma _{11}}(t)}$ is valid within the DA.

The DA procedure is rigorous (exact) if the field undergoes only phase fluctuations, while its amplitude and therefore intensity are constant \cite{ref22}. It has been shown, however, both analytically \cite{ref22,ref23,ref24} and numerically \cite{ref21}, that even in the presence of intensity fluctuations, the DA is valid as long as the Rabi frequency is not larger than the dominant relaxation constant of the excited state, which in our problem is ${\Gamma}$. In fact, the numerical justification \cite{ref21} has been derived in the context of the type of fluctuations present in the FEL radiation of FLASH. Given the values of the parameters for the He(2s2p) AI resonance, for peak intensities up to $2\times{10^{14}}{W/cm^{2}}$ the problem is within the validity of the DA approximation. And as mere inspection of the form of the above equations shows, in that case, the only effect of the field  fluctuations is the addition of the laser bandwidth to the other off-diagonal relaxation parameters. One might then be tempted to infer that the resulting profile would simply reflect this additional width. As we will see, however, the situation turns out to be much more intricate.

We begin by considering an example with conditions typical to FEL-FLASH radiation. In addition to the atomic parameters given above, we need values for the peak intensity, the pulse duration and the laser bandwidth.
Adopting, for the purposes of this quantitative illustration, values typical to experiments at FLASH, we obtain the results depicted in Fig.1. None of the four profiles, even the one for the lowest intensity, resembles the usual textbook profile $P$, which is shown in the inset of Fig.1. An attempt to fit even the lowest intensity (blue) curve with the standard parameters $q$ and $\varepsilon$, would lead to a totally irrelevant value of $q$. The most glaring difference, noticeable almost visually, is in the ratio between the value at the peak over that at the background. While in the inset it is about 9, it is of the order of 2 or less in all of the profiles in Fig.1

\begin{figure}[t]
	\centering
		\includegraphics[width=12cm]{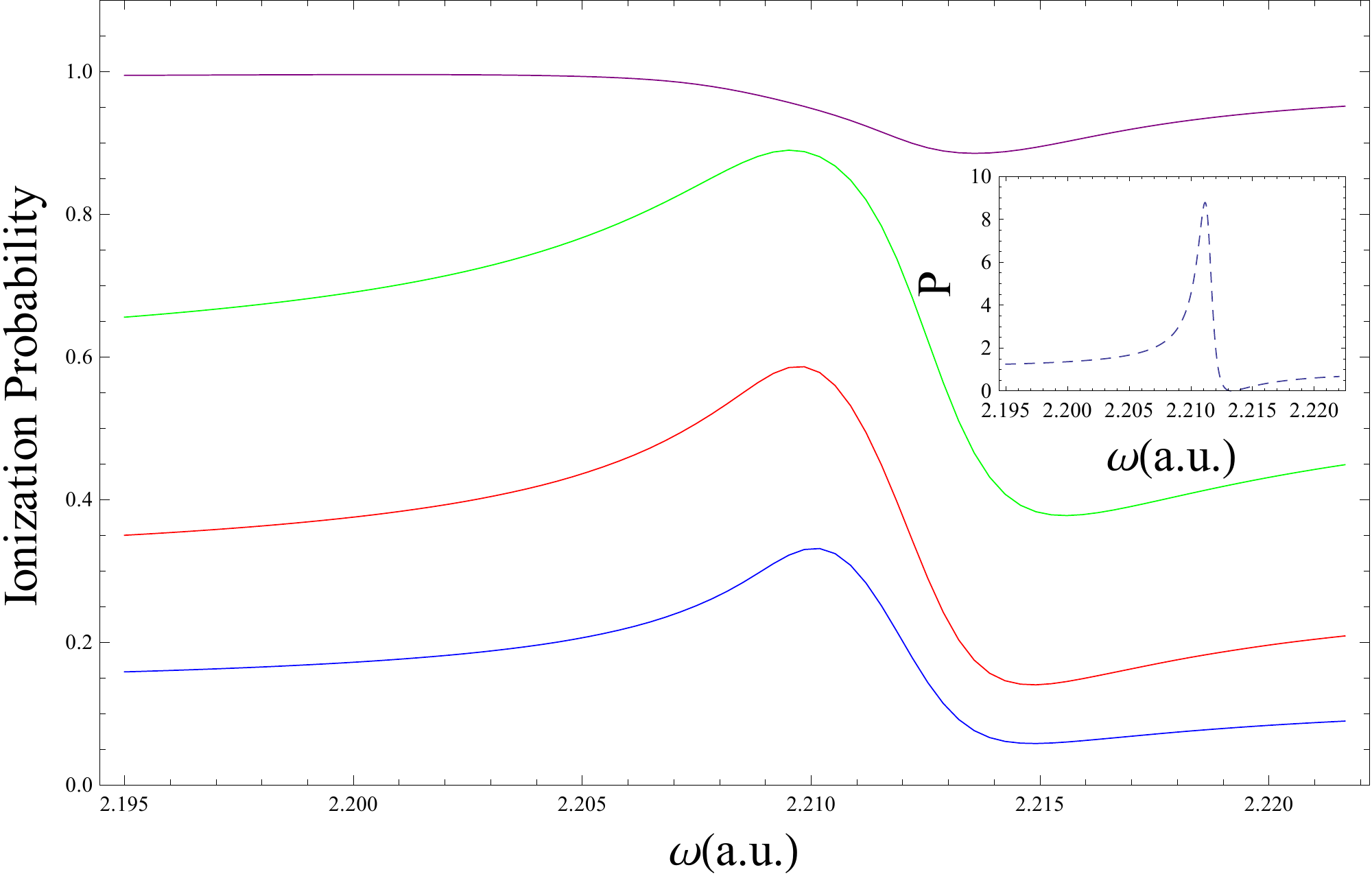}
		\rule{30em}{0.3pt}
	\caption[Fig1]{Probability of ionization as a function of the driving frequency for various intensities with pulse duration $T = 50fs$ and laser bandwidth ${\gamma _L} = 0.0018 a.u.$ Blue Line: ${I_0} = 2 \times {10^{13}}W/c{m^2}$, Red Line: ${I_0} = 5 \times {10^{13}}W/c{m^2}$, Green Line: ${I_0} = {10^{14}}W/c{m^2}$ and Purple Line: ${I_0} = 5 \times {10^{14}}W/c{m^2}$. In the inset, we show the standard Fano profile  for $q=-2.75$.}
\end{figure}

A brief parenthesis is in order at this point. Since for experimental reasons, instead of the ion or electron profile, often it is the photon transmission spectrum in terms of Beer's law that is measured, in Fig.2 we show the transmission spectra corresponding to the parameters of Fig.1. The observed signal is now seen to decrease with increasing intensity, simply because in Beer's law the transmission through the medium is divided by the incoming radiation. Nevertheless, there is a one to one 
correspondence between ion (electron) signal and transmission. The distortion of the profile is obvious in both observations. Having settled the equivalence between 
ion and transmission signals, we will hereafter confine our discussion to ion spectra.

\begin{figure}[t]
	\centering
		\includegraphics[width=12cm]{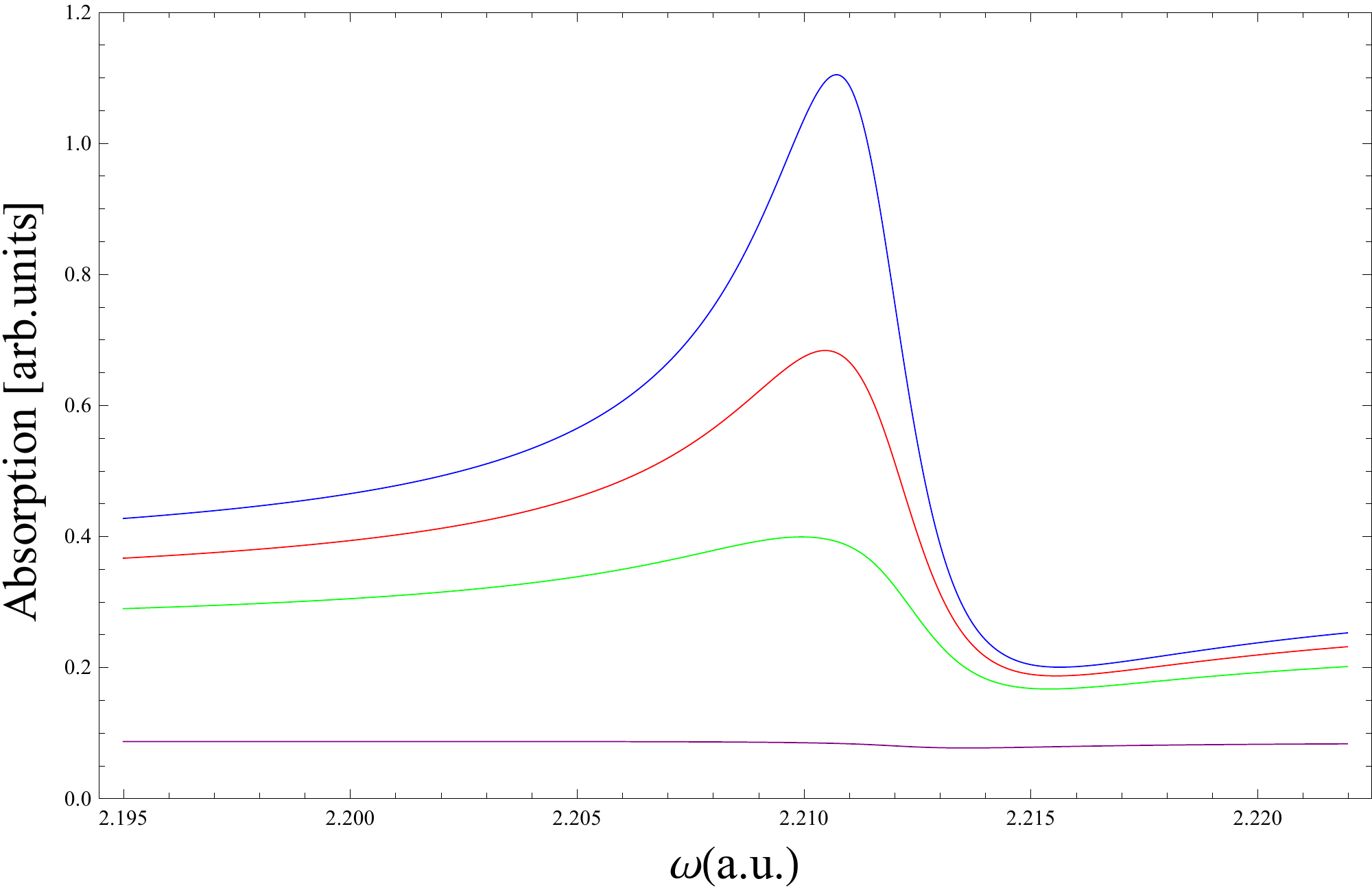}
		\rule{30em}{0.3pt}
	\caption[Fig2]{Absorption Spectrum as a function of the driving frequency for various intensities with pulse duration $T = 50fs$ and laser bandwidth ${\gamma _L} = 0.0018 a.u.$ Blue Line: ${I_0} = 2 \times {10^{13}}W/c{m^2}$, Red Line: ${I_0} = 5 \times {10^{13}}W/c{m^2}$, Green Line: ${I_0} = {10^{14}}W/c{m^2}$ and Purple Line: ${I_0} = 5 \times {10^{14}}W/c{m^2}$.}
\end{figure}

The above comparison of electron/ion spectra with their transmission counterpart, highlights one more new aspect
of AI spectra under FEL radiation. In addition to the ionization of the neutral, the radiation can ionize the $ {He(1s)}^{+}$ ions produced from autoionization. This  process involves the absorption  of one additional photon. If it is electrons or He ions that are counted, 
the resulting $\alpha$-particles do not influence the observation. But in transmission, those additional photon absorptions do contribute to the counting. The calculation must therefore include that additional channel of photon absorption, for which the cross section is ${1.2}\times{10^{-18}}{cm}^{2}$; about the same as the one for the single-photon ionization of the neutral, at the smooth part of the continuum away from the resonance.
For the sake of completeness, we have included that additional channel in our calculations. For the intensities and pulse duration in Fig. 1, its effect is barely noticeable. It could,however, make a noticeable difference for higher intensity and/or longer pulse duration. Obviously, it does not play any significant 
role under synchrotron radiation.  

From half a century or so of laser spectroscopy, we know that driving a resonant transition strongly, we
should expect a distortion of the excitation profile. For bound-bound transitions, with negligible Doppler and collisional broadening, strong driving implies a Rabi frequency larger than the dominant relaxation rate \cite{ref23,ref24}. Since for bound-bound transitions the profile tends to be Lorentzian, typically the relaxation is reflected in the width at half maximum of the profile. Even in the context 
of XUV or shorter wavelength radiation, where Auger decay may be the dominant relaxation, the profile is Lorentzian \cite{ref21}. In autoionization with a relatively small $q$ parameter, however, the AI width $\Gamma$ does not correspond to the width of the AI at half maximum. Even 
the notion of half maximum is not obvious in that case.
Let us, nevertheless, agree here to define the maximum, in relation to the background for large $\varepsilon$,
which, in the inset of Fig.1, leads to a value for $P$ equal to one. Sidestepping straightforward mathematical details here, let us note that for the case of He(2s2p) with $q=-2.75$, the width $\Gamma$ of the profile is to be found slightly above the half maximum, in the sense defined above. 

For Lorentzian profiles, strong driving usually leads to what is referred to as power broadening, which means that the profile tends to become "fatter" \cite{ref25}.
In the limit of Rabi frequency much larger then the relaxation constant, we have an AC Stark splitting, which is observable if one of the resonant (usually the upper) states, is probed by a weak transition to another state.
These issues have been formulated and discussed rather extensively even for AI states \cite{ref11,ref12,ref13}. The effect of field fluctuations, including intensity fluctuations, for strong coupling in bound-bound transitions, has been studied in exhaustive detail \cite{ref23,ref24}, but not in the case of AI resonances; simply because sources for the strong driving in the XUV and beyond did not exist, until the recent advent of the short wavelength FEL. A step in that direction has been reported in \cite{ref21}, in which the behavior of an Auger resonance driven strongly by a field with intensity fluctuations, such as those of the FEL, has been studied in great detail. To the best of our knowledge, strong driving of an AI state, such as the He(2s2p), under FEL radiation has not been observed, but this is probably a matter of short time. And the theory of an asymmetric AI state driven strongly by a field with intensity fluctuations, such as that of current FEL's turns out to be quite challenging.

Returning now to the case of driving below the strong Rabi boarder-line, we have already shown unexpected distortion of the profile, as documented by the three lowest intensity curves in Fig.1. We can safely rule out any visible  contribution from power broadening. However, although the intensity may be below the strong field, the amount of ionization by the end of the pulse can still be substantial. The point to be stressed, in this connection, is that for a pulsed source of significant intensity, pulse duration and peak intensity 
cannot be viewed independently. For example, a pulse of
50 fs duration may sound short and a peak intensity of
${10^{14}}W/c{m^2}$ may be below the strong coupling value. Put the two together, and substantial ionization
occurs, which means that for that intensity, a pulse of 
50 fs duration is a long pulse. In a time dependent situation, such as the one embodied in the density matrix equations, the amount of ionization is not simply proportional to time, as in Fermi's golden rule. As a result the ionization on resonance increases differently than it does in the wings of the profile; hence the profile distortion. One might conjecture that the distortion due to pulse duration would be minimized, or even eliminated, by decreasing the pulse duration. However, the Fourier bandwidth is lurking in that process and eventually  distortion due to the increased Fourier bandwidth, inherently included in the time-dependent calculation, begins setting in. Lest the reader be concerned with the particular temporal shape of the pulse, which has been a Gaussian in our calculations, long experience with similar calculations, including those in the present context, has shown that such details have minimal quantitative effect on the main features of the problem.

The reader may have noticed that one of the curves in Figs.1 and 2 has been obtained with an intensity larger, by a factor of 2.5, than the strong coupling limit, in which case the DA has began to lose its validity. From related calculations for an Auger resonance \cite{ref21}, we know that the error is not sufficiently large to alter the main features. A qualitative argument drawn for the case of bound transitions, can be fairly convincing here. Since the Rabi frequency is proportional to the field, a factor of 2.5 in intensity, entails a factor of about 1.58 increase in the Rabi frequency. This would imply a factor of about 1.58 in the apparent width of the resonance. Yet the profile for that intensity, bears no resemblance whatsoever to a Fano resonance. Actually it looks like what is usually referred to as "window resonance", exhibiting only a shallow minimum; a rather dramatic illustration of the interplay between pulse duration and intensity, brought about by a mere factor of 2.5 of increase in intensity. Obviously the reason the minimum is not sharp has to do with the bandwidth of the source, as the wings of the radiation profile sample transition amplitude around the minimum.   

The picture emerging from the above results should now be clear. The excitation of an AI resonance by a pulsed source, of intensity even below the strong coupling limit,  will exhibit a profile that depends on the combination of the source parameters. Under such conditions, one cannot expect to observe the textbook profile $P$. We have, moreover, demonstrated extreme sensitivity to the combination of source parameters. After all, an uncertainty of a factor of 2 in the peak intensity delivered by an FEL is not that exorbitant. The positive side of our findings is that the modified AI profile can serve as a probe of the source parameters. That is because in our calculations, we have found that a particular profile is rather sensitive to the combination of parameters compatible with its shape. For example, one may have to decide which of the three main parameters, be it peak intensity, bandwidth, pulse duration, is/are known with higher accuracy, so as to extract the value of one or two of the others. Related details, will be published in a longer paper, including certain aspects of the effect of intensity fluctuations under strong driving. The latter represents work still in progress. Definitive and complete results on that issue, such as those for an Auger resonance \cite{ref21}, will be necessary when double resonance, driven strongly by a source such as the FEL, including the observation of the resulting AC Stark splitting, become experimentally feasible.  

Acknowledgements: The authors gratefully acknowledge many and ongoing discussions with Drs. Thomas Pfeifer and Christian Ott, on experimental as well as theoretical issues related to this work. We are also indebted to Dr
G. M. Nikolopoulos  for his careful reading of the manuscript and insightful comments during the work.

\end{document}